# TEMPORAL-SPATIAL REPRESENTATION LEARNING TRANSFORMER FOR EEG-BASED EMOTION RECOGNITION

*Zhe Wang*[1]    *Yongxiong Wang*[1]    *Chuanfei Hu*[2]    *Zhong Yin*[1]    *Yu Song*[3]

[1]School of Optical-Electrical and Computer Engineering, University of Shanghai for Science and Technology, China
[2]Key Laboratory of Measurement and Control of CSE Ministry of Education, Southeast University, China
[3]Tianjin Key Laboratory for Control Theory and Applications in Complicated Systems, Tianjin University of Technology, China

**ABSTRACT**

Both the temporal dynamics and spatial correlations of Electroencephalogram (EEG), which contain discriminative emotion information, are essential for the emotion recognition. However, some redundant information within the EEG signals would degrade the performance. Specifically, the subjects reach prospective intense emotions for only a fraction of the stimulus duration. Besides, it is a challenge to extract discriminative features from the complex spatial correlations among a number of electrodes. To deal with the problems, we propose a transformer-based model to robustly capture temporal dynamics and spatial correlations of EEG. Especially, temporal feature extractors which share the weight among all the EEG channels are designed to adaptively extract dynamic context information from raw signals. Furthermore, multi-head self-attention mechanism within the transformers could adaptively localize the vital EEG fragments and emphasize the essential brain regions which contribute to the performance. To verify the effectiveness of the proposed method, we conduct the experiments on two public datasets, DEAP and MAHNOB-HCI. The results demonstrate that the proposed method achieves outstanding performance on arousal and valence classification.

## 1. INTRODUCTION

Emotion recognition has attracted more attention in the last decade, because it is essential to the rapidly developed human-computer interaction (HCI). Physiological signal, facial expression, and speech are the common modalities in the emotion recognition. Among the physiological signals, EEG could measure the amygdala activities which are closely related to the emotions [1]. Besides, EEG-based emotion recognition can be widely used in service robot [2], medical diagnose [3], game entertainment [4], and so on.

The conventional methods mainly extract handcrafted features of EEG, and then the features are fed to the deep learning model to recognize the emotional state. Rayatdoost et al. [5] transform the Power Spectral Density (PSD) features into topographic maps and feed maps into the Convolutional Neural Network (CNN). Gao et al. [6] design a dense CNN to learn the Differential Entropy (DE) features. Khare et al. [7] exploit the smoothed pseudo-Wigner–Ville distribution to obtain the temporal-spectral representation which is learned by the CNN. These methods motivate the development of EEG-based emotion recognition, but the handcrafted features would potentially lose the detailed EEG fluctuation information which is beneficial to the performance. For example, β band (13-30 Hz) and γ band (30-47 Hz) are the common choices in the PSD and DE extraction. The fairly long band-width would cause the dynamic temporal information loss.

To address this problem, some researchers focus on the end-to-end model to deal with the raw EEG signals. Alhagry et al. [8] utilize a Long Short-Term Memory (LSTM) network to capture temporal dynamics. Ding et al. [9] propose a 1D-CNN based model which combines the spatial-temporal feature extraction and classification. Bethge et al. [10] adopt DeepConvNet architecture as the private encoders to extract latent features which are fed to a shared classifier. These methods could extract discriminative EEG features by capturing the temporal dynamics and spatial correlations among the electrodes.

Raw EEG signal provides detailed information which is beneficial to discriminate the emotion, but the high-dimensional input would cause the feature redundancy. One

This work is sponsored by Natural Science Foundation of Shanghai under Grant No. 22ZR1443700. Yongxiong Wang is the corresponding author. (wyxiong@usst.edu.cn)

reason might be that only some specific plots in the stimuli materials can induce subject's prospective intense emotions rather than the whole duration [11]. In addition, extract discriminative features through the complex spatial correlations among a number of electrodes is also a challenge problem in EEG feature learning.

Recently, transformer-based models have been proposed for different EEG tasks. The multi-head self-attention within the transformer emphasizes the contributive features so that the model could effectively capture dynamic EEG context. Phan et al. [12] propose the SleepTransformer to recognize the sleep staging, and achieve outstanding performance by intra-epoch and inter-epoch context learning. Sun et al. [13] propose a hybrid model which combines the transformer with CNN for motor imagery classification. These researches demonstrate that transformer-based model could robustly extract discriminative EEG features.

In this work, we propose a transformer-based end-to-end model, denoted as Temporal-Spatial EEG Representation Transformer (TSERT), to deal with the feature redundancy of EEG and extract discriminative features by capturing temporal dynamics and spatial correlations. The model includes two major modules: weight-shared temporal feature extractors and a hierarchical spatial encoder. Different from the handcrafted feature extraction, the temporal feature extractors are weight-shared transformer encoders which adaptively emphasize the contributive representations from raw EEG signals of each channel. Next, the latent representations are fed to a hierarchical spatial encoder which robustly captures the EEG spatial correlations from the electrode level to the brain-region level. And we have preliminarily explored this module in the previous work [14]. Finally, the arousal and valence predictions are obtained by TSERT. The contributions of our work can be concluded as follows:

1) We propose a novel transformer-based architecture to learn the discriminative temporal-spatial representations. The end-to-end inputs preserve the detailed information of EEG, and multi-head self-attention could alleviate the feature redundancy.

2) The weight-shared temporal feature extractors are inspired by the fact that handcrafted feature extraction adopts the same rule among the EEG channels. And it could adaptively extract the essential information without domain knowledge.

## 2. METHODOLOGY

### 2.1. Overview of the TSERT

As shown in Fig.1, the TSERT contains two crucial parts: temporal feature extraction and hierarchical spatial learning. In the temporal feature extraction, raw EEG signals of each channel are equally divided into several slices. Next, EEG slices are considered as the patches to the transformer encoder and fed into temporal feature extractors which share the weights among all the channels. Then the latent features could be obtained by the extractors, and they are divided into different feature sets according to the region classification of the cortex. In the hierarchical spatial learning, the feature sets are separately fed to the corresponding electrode-level transformer encoders. And the obtained representations of different brain regions are served as the patches to a brain-region-level spatial encoder in order to learn the global spatial information. Finally, a sigmoid classifier is utilized for the valence and arousal prediction.

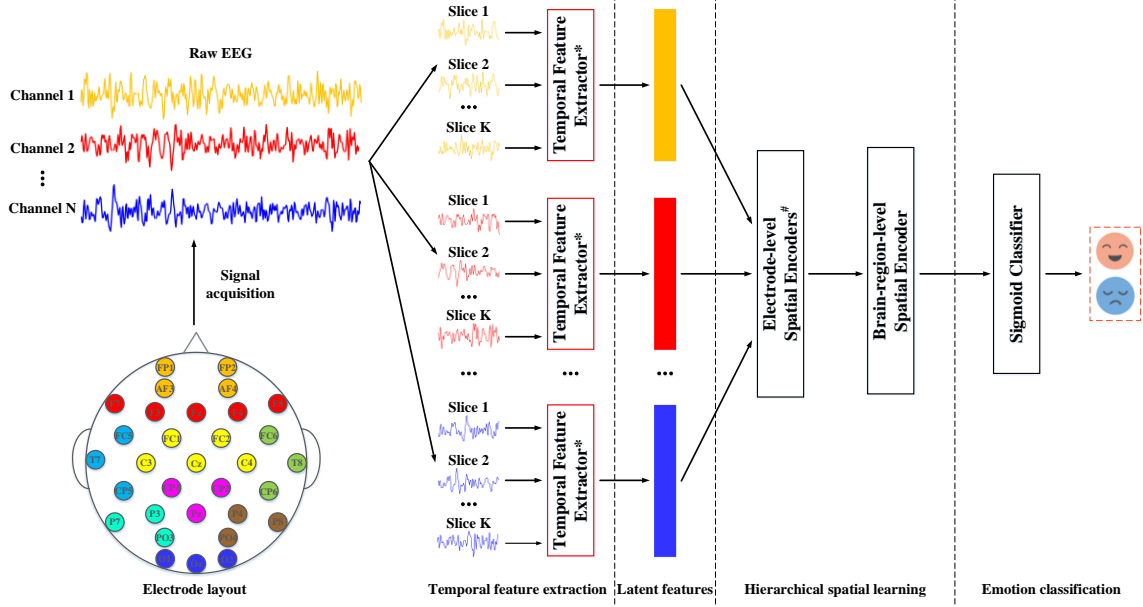

**Fig. 1.** The overview of TSERT. '*' denotes that temporal feature extractors are weight-shared among all channels. '#' denotes that latent features of different brain regions are separately fed to the corresponding electrode-level transformers. And brain regions are classified according to different brain functions, the electrodes with the same color are divided into a brain region.

## 2.2. Temporal feature extraction

Given the representation of raw EEG signals with $N$ channels $X = [X^1, X^2, ..., X^N] \in \mathbb{R}^{N \times d}$, where $d$ is signal length. The weight of temporal feature extractor is share among all the channels. In each channel, the EEG signals are divided into $K$ patches $X^n = [X_1^n, X_2^n, ..., X_K^n] \in \mathbb{R}^{K \times d'}$ $n = 1, 2, ..., N$, where patch length $d' = \frac{d}{K}$. Firstly, the electrode patches are mapped to a constant size $D_T$ using linear embedding with the weight $W_T \in \mathbb{R}^{d' \times D_T}$. Next, the class token $X_T^{cls} \in \mathbb{R}^{D_T}$ is inserted at the head of the patches to aggregate representative information. Besides, positional embedding $E_T^{pos} \in \mathbb{R}^{(K+1) \times D_T}$ is added to the patches to preserve positional information. These operations can be represented as follows:
$$Z_0 = [X_T^{cls}, X_1^n W_T, ..., X_K^n W_T] + E_T^{pos}$$
where $Z_0 \in \mathbb{R}^{(K+1) \times D_T}$ is the input of transformer encoder.

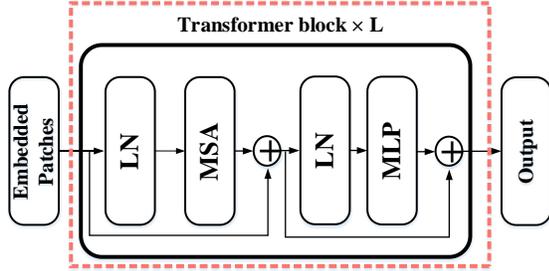

**Fig. 2.** The structure of transformer encoder

As shown in Fig.2, the transformer encoder includes: Multi-head Self-Attention (MSA), Multiple Layer Perception (MLP), and Layer Normalization (LN). Hence, the operation within the transformer encoder is shown as:
$$Z_l' = MSA(LN(Z_{l-1})) + Z_{l-1} \quad l = 1, ..., L_T$$
$$Z_l = MLP(LN(Z_l')) + Z_l' \quad l = 1, ..., L_T$$
where $MSA(\cdot)$ denotes MSA operation, and $MLP(\cdot)$ denotes MLP operation, the $L_T$ is the number of transformer blocks, and the final output is $Z_{L_T} \in \mathbb{R}^{(K+1) \times D_T}$. We take the average of $Z_{L_T}$ as the latent feature $\bar{Z}_{L_T} \in \mathbb{R}^{D_T}$ of the current channel. Finally, the obtained latent features of all the channels are concatenated as $Z_T \in \mathbb{R}^{N \times D_T}$.

## 2.3. Hierarchical spatial learning

Before the spatial learning, the latent feature $Z_T$ are divided into nine subsets according to nine brain regions (Pre-frontal, frontal, and so on) which has been preliminarily explored in our previous work [14]. In the electrode-level spatial learning, the linear embedding and positional embedding can be represented as follows:
$$Z_0^E = [X_E^{cls}, X_1 W_E, ..., X_M W_E] + E_E^{pos}$$
where $M$ is the number of the electrodes within the brain region, $Z_0^E \in \mathbb{R}^{(M+1) \times D_E}$ is the transformer input, $D_E$ is the patch dimension, $W_E \in \mathbb{R}^{D_T \times D_E}$ is the weight, $X_E^{cls} \in \mathbb{R}^{D_E}$ is the class token, and $E_E^{pos} \in \mathbb{R}^{(M+1) \times D_E}$ is the positional embedding. Similar to the aforementioned, the input of brain-region-level spatial learning could be written as:
$$Z_0^B = [X_B^{cls}, X_1 W_B, ..., X_9 W_B] + E_B^{pos}, Z_0^B \in \mathbb{R}^{(9+1) \times D_B}$$

Besides, the operation within the transformer is also similar to the temporal feature extraction. The numbers of the block in the electrode-level and brain-region-level are $L_E$ and $L_B$ respectively. And the outputs are $Z_{L_E} \in \mathbb{R}^{(M+1) \times D_E}$ and $Z_{L_B} \in \mathbb{R}^{(9+1) \times D_B}$. The output patch according to the class token $Z_{L_B}^0 \in \mathbb{R}^{D_B}$ is utilized to predict the emotion. The binary prediction of arousal or valence is obtained as follows:
$$\hat{y} = \sigma(W_O Z_{L_B}^0)$$
where $\hat{y}$ is the prediction and $\sigma(\cdot)$ denotes the sigmoid function.

## 3. EXPERIMENTS AND RESULTS

### 3.1. Dataset and pre-processing

In this work, the DEAP [15] and MAHNOB-HCI [16] are chosen as the benchmark emotion datasets. Firstly, the 32-channel EEG signals are down-sampled from 512Hz to 128 Hz. Next, a 4-45 Hz band pass filter and independent component analysis are adopted to remove the artifacts. Finally, we adopt a 6-second-long sliding window without overlap to segment the EEG data. Each segment within a trial is considered as a sample.

On the other hand, the self-assessment value of 1-4 is considered as 'low class', the low arousal (LA) and low valence (LV). And the self-assessment value of 6-9 is considered as 'high class', the high arousal (HA) and high valence (HV). Therefore, we transform the emotion recognition problem into two binary classifications (LA vs. HA, and LV vs. HV).

### 3.2. Experiment setup

In the input settings, we set the $d$ as 768 (6s × 128Hz), $K$ as 6, and $d'$ as 128. And the hyperparameters of TSERT are also necessary to determine. In the module of temporal feature extraction, the $D_T$ and $L_T$ are set as 64 and 1 respectively. In the module of hierarchical spatial learning, the $D_E$, $D_B$, $L_E$, and $L_B$ are set as 32, 64, 2, and 2 respectively. In this work, all the networks are implemented by Pytorch with a NVIDIA GeForce RTX 3090 GPU. We adopt the Adam optimizer with cosine learning decay, learning rate is $10^{-4}$, batch size is 512, and epoch is 80 with early stopping.

### 3.3. Results

To validate the effectiveness of TSERT, we design several variants. The details of these variants are as follows:
(1) SERT: Compared with TSERT, the temporal feature extractors are removed from TSERT.
(2) TERT: Compared with TSERT, the hierarchical spatial learning module is replaced by an output layer.

**Table 1.** The results of TSERT and its variants (%)

| Model | DEAP | | | | MAHNOB-HCI | | | |
|---|---|---|---|---|---|---|---|---|
| | Arousal | | Valence | | Arousal | | Valence | |
| | Acc | F1-Score | Acc | F1-Score | Acc | F1-Score | Acc | F1-Score |
| SERT | 64.98 | 63.31 | 64.48 | 62.11 | 65.02 | 62.69 | 63.80 | 62.71 |
| TERT | 65.02 | 63.56 | 63.71 | 61.98 | 64.80 | 62.67 | 63.78 | 62.55 |
| EEGNet | 61.93 | 59.14 | 61.32 | 59.57 | 62.17 | 59.53 | 61.73 | 60.78 |
| STERT | 65.83 | 62.63 | 64.33 | 61.45 | 65.39 | 63.10 | 64.88 | 63.41 |
| TSERT (PSD input) | 66.76 | 64.42 | 65.73 | 63.51 | 67.21 | 64.43 | 66.93 | 65.49 |
| **Proposed TSERT** | **68.87** | **66.76** | **67.59** | **66.44** | **70.02** | **66.62** | **69.32** | **66.39** |

(3) STERT: Different from TSERT, it firstly learns the spatial information, and then it captures the temporal context of EEG. The weights of the transformer encoders within hierarchical spatial learning are shared. And one transformer encoder is adopted to capture the temporal context among the slices.
(4) TSERT (PSD input): The raw EEG inputs are replaced by PSD features.
(5) EEGNet [17]: It is a classic end-to-end EEG representation model, and it also captures the temporal and spatial information. We reimplement it for the comparison.

In this work, we adopt leave-one-out cross-validation to evaluate the classification performance. Specifically, testing set is the EEG data from one subject, and training set is the EEG data from the others. This process will recurrent until each subject's data has been set as testing set once. And overall performance is the average of all the folds.

The experiment results are shown in Table 1, we adopt the accuracy (denoted as Acc) and F1-Score to evaluate the performance. The proposed TSERT has achieved the best performances among all the models. It achieves the accuracy of 68.87% and 67.59% at arousal and valence level on the DEAP, and 70.02% and 69.32% on the MAHNOB-HCI.

Compared with SERT and TERT, the TSERT have boosted the accuracy more than 3% on the DEAP dataset, and 5% on the MAHNOB-HCI. This could validate the effectiveness of temporal feature extraction module and hierarchical spatial learning module. Compared with EEGNet, the TSERT also achieved the accuracy improvement (about 6% on the DEAP and MAHNOB-HCI dataset). It demonstrates that the self-attention within the TSERT could obtain more discriminative representations. And the TSERT is also superior to the STERT (about 3% on the DEAP dataset, and about 5% on the MAHNOB-HCI dataset). The reason might be that raw EEG signals are non-linear and non-stationary, and it is difficult for the model to directly learn the spatial information. Nevertheless, STERT still outperforms the SERT and TERT. It indicates that learning discriminative representation from both temporal and spatial domain is better than the method only from single domain. Besides, we also compare the impact of different input forms on performance of TSERT. According to the results, we find that the performance of raw EEG signals surpasses the PSD features. It indicates that the TSERT could take full advantage of the detailed dynamic temporal information which is beneficial to discriminate the emotions.

**Table 2.** The performance comparison with recent works using cross-subject strategy

| References | Model | Acc (%) |
|---|---|---|
| Li et al. [18] (2018) | SVM | A: - V: 59.06 (D) |
| Pandey et al [19] (2019) | Deep Neural Network | A: 61.25 V: 62.50 (D) |
| Hagad et al. [20] (2021) | BiVDANN | V: 63.52 (D) |
| Yin et al. [21] (2020) | Locally-robust feature selection & LSSVM | A: 65.10 V: **67.97** (D) A: 67.43 V: **70.90** (M) |
| Zhang et al. [22] (2020) | Shared-subspace feature elimination | A: 65.21 V: 66.35 (D) A: 65.20 V: 65.37 (M) |
| Ding et al. [9] (2022) | Multi-scale 1D-CNN | A: 61.57 V: 62.33 (D) A: 60.61 V:61.27 (M) |
| Wang et al. [14] (2022) | HSLT | A: 65.75 V: 66.51 (D) A: 66.20 V: 66.63 (M) |
| Ours | TSERT | A: **68.87** V: 67.59 (D) A: **70.02** V: 69.32 (M) |

Note: A and V denote the accuracy of arousal and valence; D denotes the performance on DEAP dataset; M denotes the performance on MAHNOB-HCI dataset; BiVDANN denotes Bi-lateral Variational Domain Adversarial Neural Network; HSLT denotes Hierarchical Spatial Learning Transformer.

Furthermore, we compare the results with the recent works using cross-subject strategy in Table 2. The results indicates that it is more difficult to design a cross-subject EEG recognition model. The reason might be the inconsistent data distribution among the different subjects. Overall, TSERT has achieved outstanding performance among the recent works. And it also outperforms the HSLT which is proposed in our previous work.

## 4. CONCLUSION

In this work, we propose a transformer-based end-to-end model, TSERT, to learn the discriminative representations from both temporal and spatial domain for EEG-based emotion recognition. Overall, the TSERT has achieved the outstanding performance on two public datasets. Especially, weight-shared temporal feature extractors outperform the handcrafted feature extraction. It could be attributed to the superior ability of multi-head self-attention in sequence learning. Specifically, the shared temporal feature extractor could adaptively represent the contributive temporal context, and alleviate feature redundancy. In the future work, we will adopt transfer learning method to optimize the TSERT in the cross-subject emotion recognition.

## 5. REFERENCES

[1] H. Gunes and B. Schuller, "Categorical and dimensional affect analysis in continuous input: Current trends and future directions," *Image and Vision Computing*, vol. 31, no. 2, pp. 120–136, February 2013.

[2] M. Val-Calvo, J. R. Alvarez-Sanchez, J. M. Ferrandez-Vicente, and E. Fernandez, "Affective Robot Story-Telling Human-Robot Interaction: Exploratory Real-Time Emotion Estimation Analysis Using Facial Expressions and Physiological Signals," *IEEE Access*, vol. 8, pp. 134051-134066, July 2020.

[3] H. Huang, Q. Xie, J. Pan, Y. He, Z. Wen, R. Yu, and Y. Li, "An EEG-Based Brain Computer Interface for Emotion Recognition and Its Application in Patients with Disorder of Consciousness," *IEEE Transaction on Affective Computing*, vol. 12, no. 4, pp. 832-842, October 2021.

[4] D. J. Diaz-Romero, A. M. R. Rincón, A. Miguel-Cruz, N. Yee and E. Stroulia, "Recognizing Emotional States with Wearables While Playing a Serious Game," *IEEE Transaction on Instrumentation and Measurement*, vol. 70, pp. 2506312, February 2021.

[5] S. Rayatdoost and M. Soleymani, "Cross-corpus EEG-based emotion recognition," in *2018 IEEE International workshop on Machine Learning for Signal Processing (MLSP)*, Aalborg, Denmark, 2018, pp. 1–6.

[6] Z. Gao, X. Wang, Y. Yang, Y. Li, K. Ma, and G. Chen, "A Channel-Fused Dense Convolutional Network for EEG-Based Emotion Recognition," *IEEE Transactions on Cognitive and Developmental Systems*, vol. 13, no. 4, pp. 945-954, December 2021.

[7] S. K. Khare and V. Bajaj, "Time–Frequency Representation and Convolutional Neural Network-Based Emotion Recognition," *IEEE Transactions on Neural Networks and Learning Systems*, vol. 32, no. 7, pp. 2901-2909, July 2021.

[8] S. Alhagry, A. Aly, and R. A., "Emotion Recognition based on EEG using LSTM Recurrent Neural Network," *International Journal of Advanced Computer Science and Applications*, vol. 8, no. 10, pp. 355-358, August 2017.

[9] Y. Ding, N. Robinson, S. Zhang, Q. Zeng, and C. Guan, "TSception: Capturing Temporal Dynamics and Spatial Asymmetry from EEG for Emotion Recognition", *IEEE Transaction on Affective Computing*, April 2022, early access.

[10] D. Bethge, P. Hallgarten, T. Grosse-Puppendahl, M. Kari, R. Mikut, A. Schmidt, and O. Özdenizci, "Domain-Invariant Representation Learning from EEG with Private Encoders", in *2022 IEEE International Conference on Acoustics, Speech and Signal Processing (ICASSP)*, Singapore, Singapore, 2022, pp. 21928484.

[11] Y. Zhang, H. Liu, D. Zhang, X. Chen, T. Qin, and Q. Zheng, "EEG-based Emotion Recognition with Emotion Localization via Hierarchical Self-Attention," *IEEE Transaction on Affective Computing*, January 2022, early access.

[12] H. Phan, K. Mikkelsen, O. Y. Chén, P. Koch, A. Mertins, and M. D. Vos, "SleepTransformer: Automatic Sleep Staging With Interpretability and Uncertainty Quantification," *IEEE Transaction on Biomedical Engineering*, vol. 69, no. 8, pp. 2456-2467, August 2022.

[13] J. Sun, J. Xie, and H. Zhou, "EEG Classification with Transformer-Based Models," in *IEEE 3rd Global Conference on Life Sciences and Technologies (LifeTech)*, Nara, Japan, 2021, pp. 92-93.

[14] Z. Wang, Y. Wang, C. Hu, Z. Yin, and Y. Song, "Transformers for EEG-Based Emotion Recognition: A Hierarchical Spatial Information Learning Model," *IEEE Sensors Journal*, vol. 22, no. 5, pp. 4359-4368, January 2022.

[15] S. Koelstra, C. Muhl, M. Soleymani, J. S. Lee, A. Yazdani, T. Ebrahimi, T. Pun, A. Nijholt, and I. Patras, "DEAP: A database for emotion analysis using physiological signals," *IEEE Transaction on Affective Computing*, vol. 3, no. 1, pp. 18–31, January 2012.

[16] M. Soleymani, J. Lichtenauer, T. Pun and M. Pantic, "A Multimodal Database for Affect Recognition and Implicit Tagging," *IEEE Transaction on Affective Computing*, vol. 3, no. 1, pp. 42-55, January 2012.

[17] V. J. Lawhern, A. J. Solon, N. R. Waytowich, S.M. Gordon, C. P. Hung, and B. J. Lance, "EEGNet: a compact convolutional neural network for EEG-based brain–computer interfaces," *Journal of Neural Engineering*, vol. 15, no. 5, pp. 056013, July 2018.

[18] X. Li, D. Song, P. Zhang, Y. Hou and B. Hu, "Exploring EEG features in cross-subject emotion recognition," *Frontiers in Neuroscience*, vol. 12, pp. 162, March 2018.

[19] P. Pandey and K. R. Seeja, "Subject independent emotion recognition from EEG using VMD and deep learning," *Journal of King Saud University - Computer and Information Sciences*, vol. 43, no. 5, pp. 1730-1738, May 2022.

[20] J. L. Hagad, T. Kimura, K. Fukui, and M. Numao, "Learning Subject-Generalized Topographical EEG Embeddings Using Deep Variational Autoencoders and Domain-Adversarial Regularization," *Sensors*, vol. 21, no. 5, pp. 1792. March 2021.

[21] Z. Yin, L. Liu, J. Chen, B. Zhao, and Y. Wang, "Locally robust EEG feature selection for individual-independent emotion recognition," *Expert Systems with Applications*, vol. 162, pp. 113768. December 2020.

[22] W. Zhang, Z. Yin, Z. Sun, Y. Tian, and Y. Wang, "Selecting transferrable neurophysiological features for inter-individual emotion recognition via a shared-subspace feature elimination approach," *Computers in Biology and Medicine*, vol. 123, pp. 103875, August 2020.